\documentclass[a4paper,twocolumn,fleqn]{article}
\usepackage{helvet}
\usepackage[dvipdfm]{graphicx}     % dvipdfm or dvipdfmx
\usepackage{jpfr}   
\newcommand{\be}{\begin{equation}}
\newcommand{\ee}{\end{equation}}
\newcommand{\bea}{\begin{eqnarray*}}
\newcommand{\eea}{\end{eqnarray*}}
\newcommand{\beaa}{\begin{eqnarray}}
\newcommand{\eeaa}{\end{eqnarray}}

\begin{document}      

\title{Coulomb collisional relaxation process of ion beams in magnetized plasmas}

\author
{Y.~Nishimura 
}
\affiliation{Plasma and Space Science Center, National Cheng Kung University, Tainan 70101, Taiwan}

\begin{abstract}
An orbit following code is developed to calculate 
ion beam trajectories in magnetized plasmas.
The equation of motion (the Newton's equation) is solved 
including the Lorentz force term and Coulomb collisional relaxation term.
Furthermore, a new algorithm is introduced
by applying perturbation method
regarding the collision term as a small term.
The reduction of computation time is suggested.
\end{abstract} 

\maketitle

\section{Introduction}
An orbit following calculation of charged particles is one of the
classic problems.
It is straightforward, but remains to be an important tool for studying
particle confinement in laboratory plasmas. 
In high beta magnetic 
confinement devices, Larmor radius
of energetic particles can be comparable to the characteristic
scale length of the experiment
(for example $\alpha$ particles in burning plasmas).
Guiding center approximation fails in the latter cases.\cite{sol96,mik97}

In this work, the equation of motion is solved 
incorporating Lorentz force term and Coulomb collisional relaxation term.
Since solving the Lorentz force term requires much shorter time step
compared to the guiding center calculation,\cite{mor66,lit83,nis97,nis08} 
computational efficiency is the key.
We introduce a new algorithm to calculate ion beam trajectories 
in magnetized plasma by applying perturbation method regarding 
the Coulomb collisional relaxation term as a small perturbation.

We start our analysis from studying the ion orbital behavior 
(with and without collision effects) in a simple geometry
where the magnetic field is axis-symmetric. 
In this paper, we employ a theta pinch plasma\cite{mor69,ish84} in a 
two dimensional system at a plasma equilibrium. 
The orbit following calculation can be useful in studying
suppression of
tilting instabilities\cite{hor90} and rotational instabilities\cite{ish84}
by the ion beams.

In Sec.\ \ref{s2}, the basic computation model is described.
The orbit following
calculation is discussed in Sec.\ \ref{s3}. 
Section~\ref{s4} presents
the perturbation method.
We summarize this work in Sec.~\ref{s5}.

\section{Equation of motion}
\label{s2}
In this section, the equation of motion is described.
Ion beam equation in the MKS unit is given
by\cite{miy86}
\beaa
m {d {\bf v} \over dt}
&=& q {\bf v} \times {\bf B}  \nonumber \\
&-& {q^2 {\bf v} \over 4 \pi \varepsilon_0^2 v^3}
\sum^\star {\log \Lambda {q^\star}^2 \over m_r} n^\star \Phi_1 (b^\star v),
\eeaa
\be
{d {\bf x} \over  dt} =  {\bf v}.
\ee
Here we recapitulate Ref.\cite{miy86}
as precise as possible (including the notations),
for the transparency of the work.
The first and the second term of Eq.(1) are
the Lorentz force term 
(we assume the electric field to be zero)
and Coulomb collisional relaxation term, respectively.
The second term reflects the momentum change of the test particle
per unit time.\cite{miy86}
Here, $m$ and $q$ are the mass and the charge of the beam ions.
The magnetic field is given by ${\bf B}$ while the ion beam positions
and velocities are given by ${\bf x}$ and ${\bf v}$, respectively.
The vacuum permittivity is given by $ \varepsilon_0$, 
and the Coulomb logarithm (see appendix) is given by $\Lambda$.
All the variables with the superscript 
$\star$ signify that of the background plasma species 
(the ions and the electrons).
Here, $m_r = m m^\star/ (m + m^\star) $ is the reduced mass.
The function $\Phi_1 $ represents the Gaussian velocity
distribution of the background plasma (see appendix),
where $T^\star$ is the background plasma temperature
and $b^\star = (m^\star / 2 q^\star T^\star)^{1/2}$.

Equations (1) and (2) are solved in a Cartesian coordinate 
$x,y$, and $z$ 
using a fourth order Runge-Kutta-Gill method.\cite{kaw89}
Equations (1) and (2) holds for ion orbital behavior in
three dimensional magnetized plasmas in general.
In this paper, as an initial application, 
a rigid roter profile of theta pinch plasma\cite{mor69,ish84}
is employed for the two dimensional magnetic field model.
Denoting $r = (x^2 + y^2)^{1/2} $, 
the magnetic field is given by
\be
{\bf B} = B_0 \tanh{ \left[ \kappa \left( 2 r^2 / r_s^2 - 1 \right) \right] } {\bf z}.
\ee
where the background density is given by
\be
n^\star = n_0 {sech}^2 { \left[ \kappa \left( 2 r^2 / r_s^2 - 1 \right) \right] },
\ee
where $\kappa$ is a constant and $r_s$ is the radius at the separatrix.\cite{tus88}
Equations (3) and (4) are in plasma equilibrium.\cite{mor69}
Correspondingly, the magnetic flux $\psi (r) = \int^r_0 B (r') r' dr' $ is given by
\beaa
\psi (r) &=& \frac{B_0 r_s^2}{4 \kappa} 
\log{ \left[ \cosh{ \left[ \kappa \left( 2 r^2 / r_s^2 - 1 \right) \right] } \right] } \nonumber \\
&-& \frac{B_0 r_s^2}{4 \kappa} 
\log{ \left[ \cosh{ \left( \kappa  \right) } \right] } .
\eeaa
In this paper, the angular momentum is given by\cite{miy86}
\be
P_\theta = m r^2 \dot{\theta} + q \psi (r),
\ee
where $\dot{\theta}$ is the time derivative of the angular coordinate $\theta$.
The kinetic energy is given by $E_k = m {\bf v}^2 /2$.

\section{Beam ion orbit}
\label{s3}
\indent
In this section, the ion orbit calculation 
is presented employing Eqs.(1) and (2).
We study beam ion (energetic particle) 
behavior whose temperature is much larger 
than that of the background thermal plasma. 

Figure 1 shows the particle orbits in a Cartesian coordinate 
in the absence of Coulomb collisions.
The magnetic configuration
reflects that of the FRC injection experiment (FIX) parameter
in the confinement chamber;\cite{shi93}
in Eq.(3), the magnetic field strength $B_0$ is $0.05 (T)$ 
and the separatrix radius $r_s$ is given by $0.2 (m)$
and thus the magnetic null is at $r_n = 0.141 (m)$.
The wall radius is set at $r_w = 0.4 (m)$.
In Eq.(3), we set $\kappa = 0.6.$\cite{mor69}
The beam ion species is Hydrogen. Throughout this paper,
we assume that the neutral beams are ionized
at $x=-0.136 (m)$ and $y=0.147 (m)$ which is on the separatrix
(the initial position of the beam ion calculation is given there).

In Fig.1(a), the beam ion temperature is given by $T_b = 50 ( eV)$
[followed the ion orbit for $20 (\mu s)$], 
while in Fig.1(b), the beam ion temperature is given by $T_b = 4000 (eV)$ 
[followed the ion orbit for $2 (\mu s)$].
Naturally, the Fig.1(b) case has a larger Lamor radius. 
As one can see, the direction of the Larmor precession changes
when the trajectory crosses the magnetic null point "$r_n $".
This is referred to as {\it meandering motion}.\cite{hor90}
Since the magnitude of the magnetic field inside $r_n$ is weaker 
than the outside, the Larmor radius is slightly
larger inside the separatrix 
[see Fig.1(c) where the magnetic field strength
and the density profile are depicted].
As shown above the motion is periodic 
which can be understood by the Noether's theorem (a canonical
variable conjugate to a constant momentum undergoes periodic motion).
The conservation of kinetic energy and the angular momentum is verified
for the calculation in Fig.1. With a single precision, the momentum (energy)
conserves at the accuracy of $6.3 \times 10^{-5} \%$  ($5.5 \times 10^{-4} \%$)
of the absolute value after following the orbit for $10000$ steps.
Here, the time step in the calculation is given by one percent of 
$2 \pi / \Omega_c$ where $\Omega_c$ is the beam ion's cyclotron frequency.
\begin{figure}
 \centering
 \includegraphics[height=7.5cm,angle=+00] {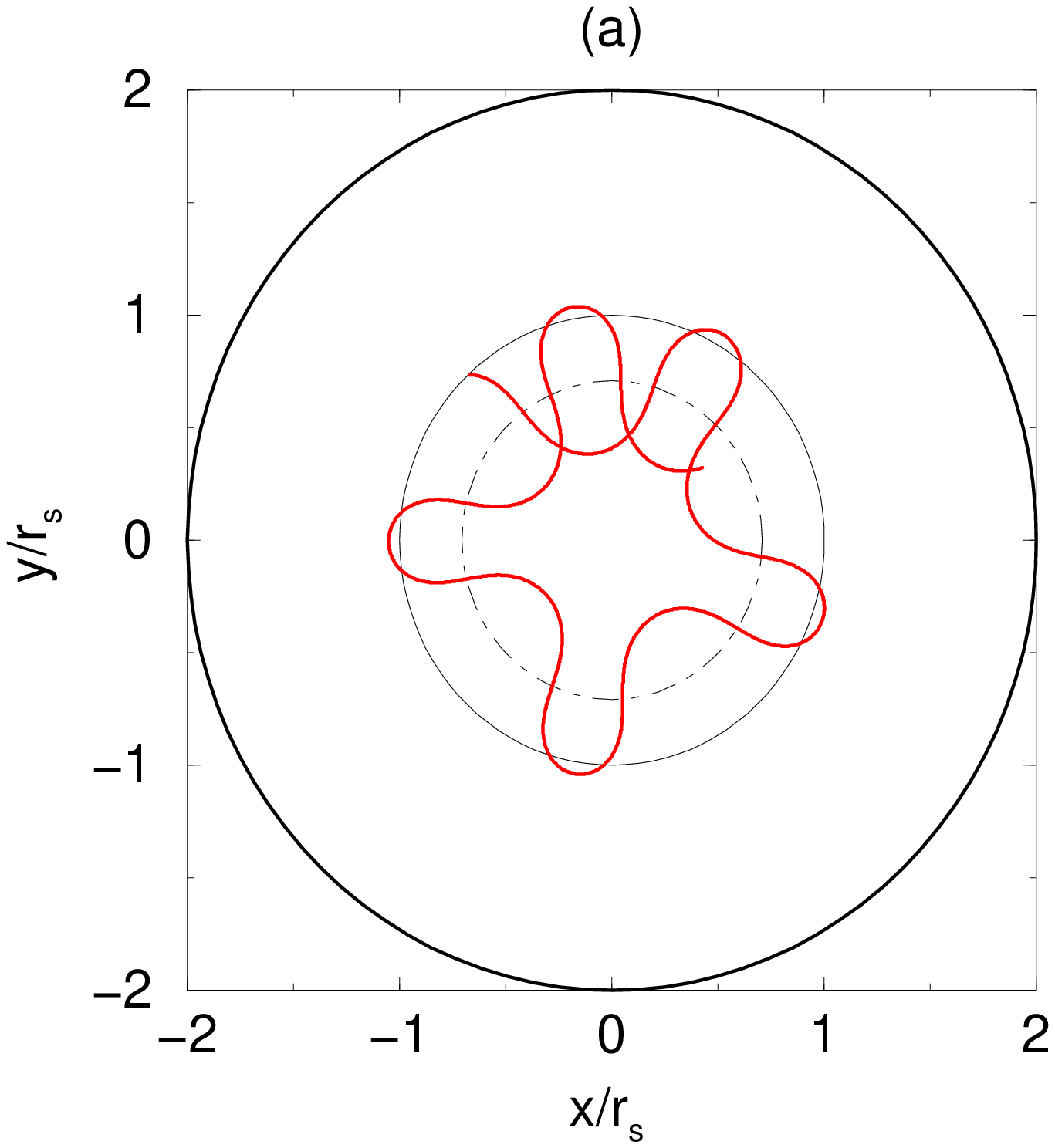}
 \includegraphics[height=7.5cm,angle=+00] {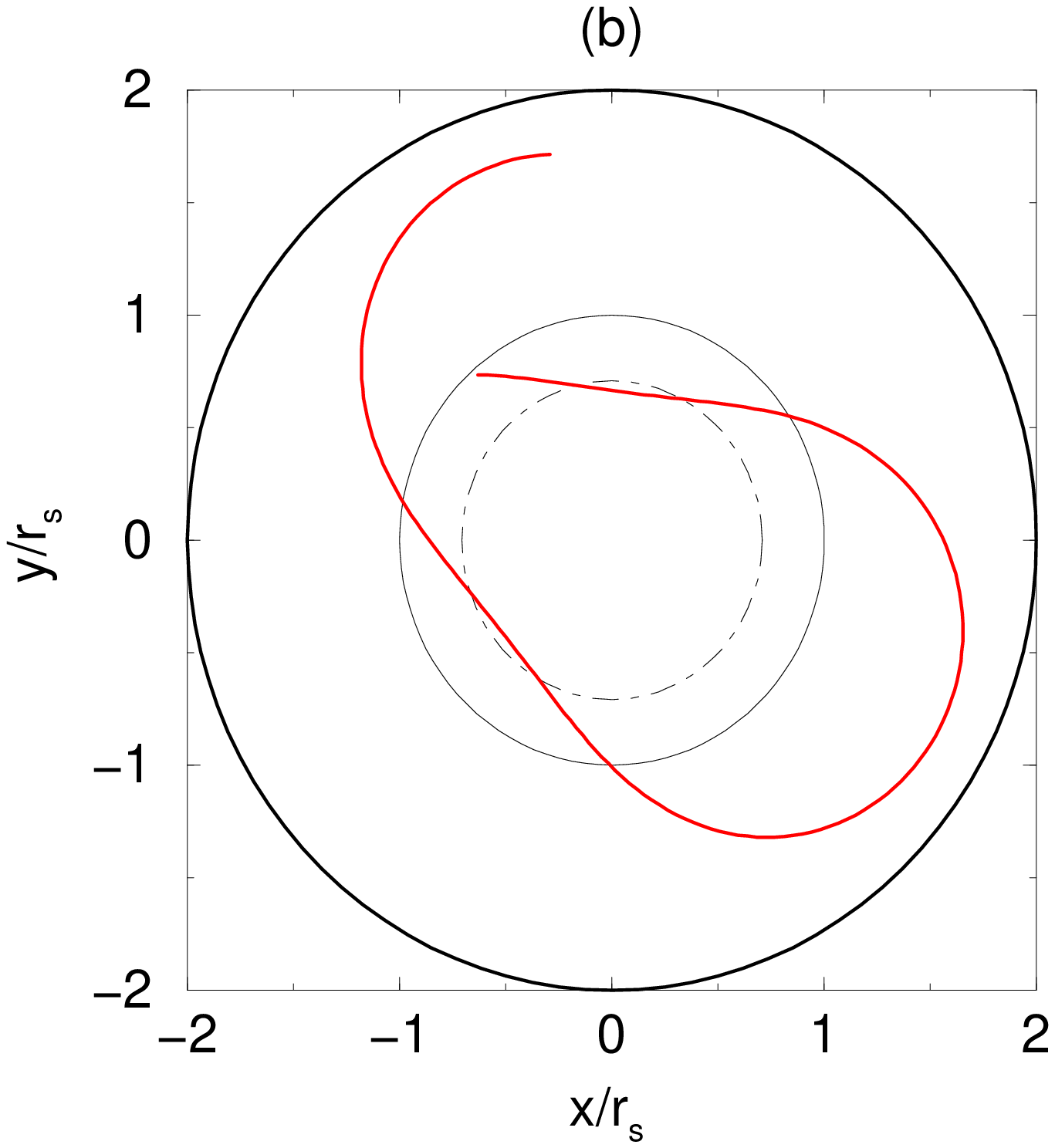}
 \includegraphics[height=7.5cm,angle=+00] {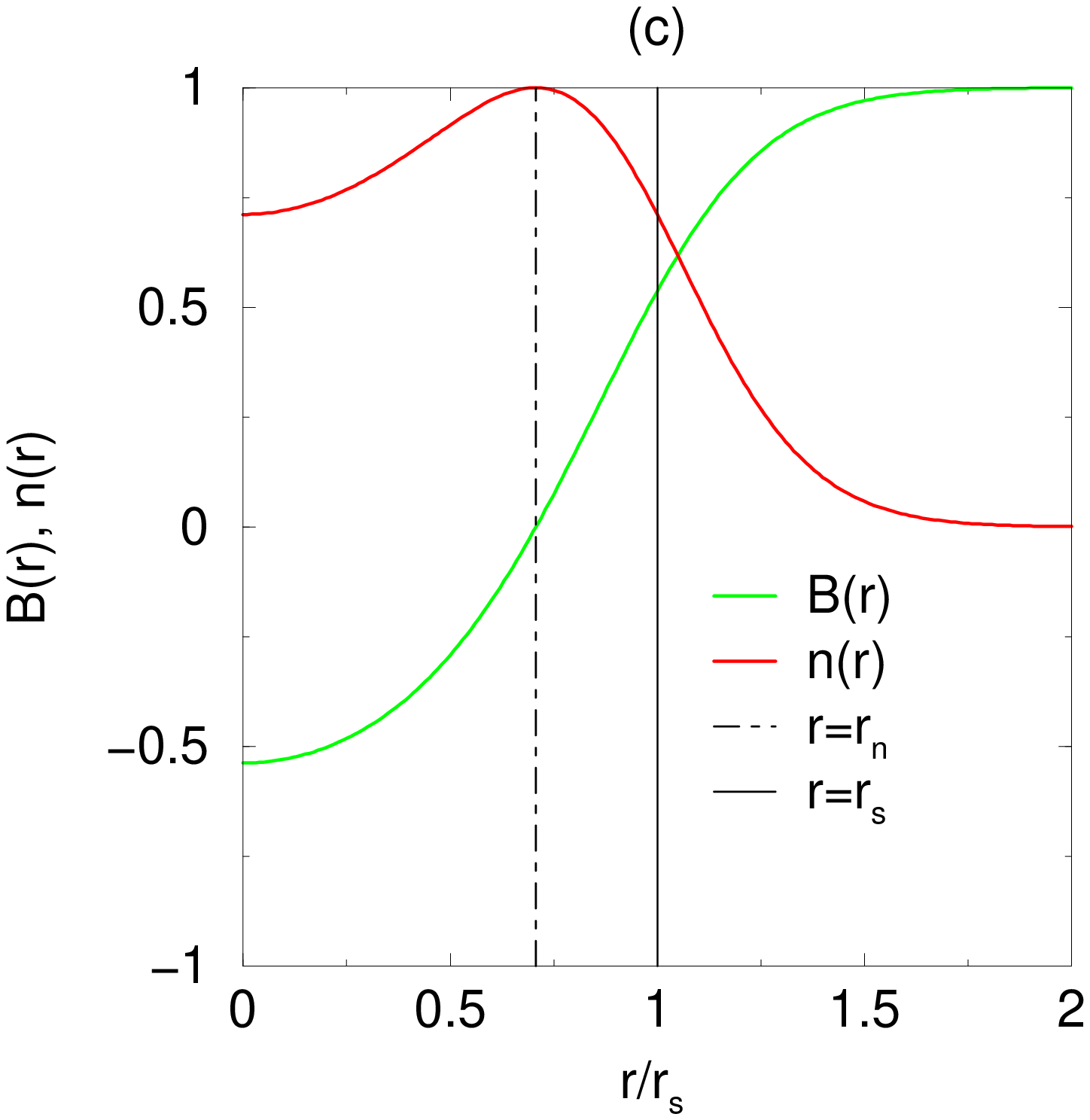}
\caption{Orbital behavior of (a) 50 (eV) (b) 4000 (eV) Hydrogen ions.
The collision term is turned off in Eq.(1).
The Larmor precession changes its direction at the magnetic null point (dashed circle).
Unit length is normalized by $r_s$ (solid circle).
(c) The magnetic field strength (green curve)
and the density profile (red curve) are depicted. 
The location of the separatrix and
the magnetic null point are suggested.}
\label{fig1}
\end{figure}

Figures 2 and 3 show the particle orbits
in the presence of Coulomb collisions.
The collision effect is dominated by electrons (see appendix).
The background electron density and temperature is given by
$n_0 = 5 \times 10^{19} (m^{-3})$ and $T_e = 20 (eV)$.
In Fig.2, the beam ion temperature is given by $T_b = 100 (eV)$.
In Fig.3, $T_b = 2000 (eV)$.
\begin{figure}
  \centering
 \includegraphics[height=7.5cm,angle=+00] {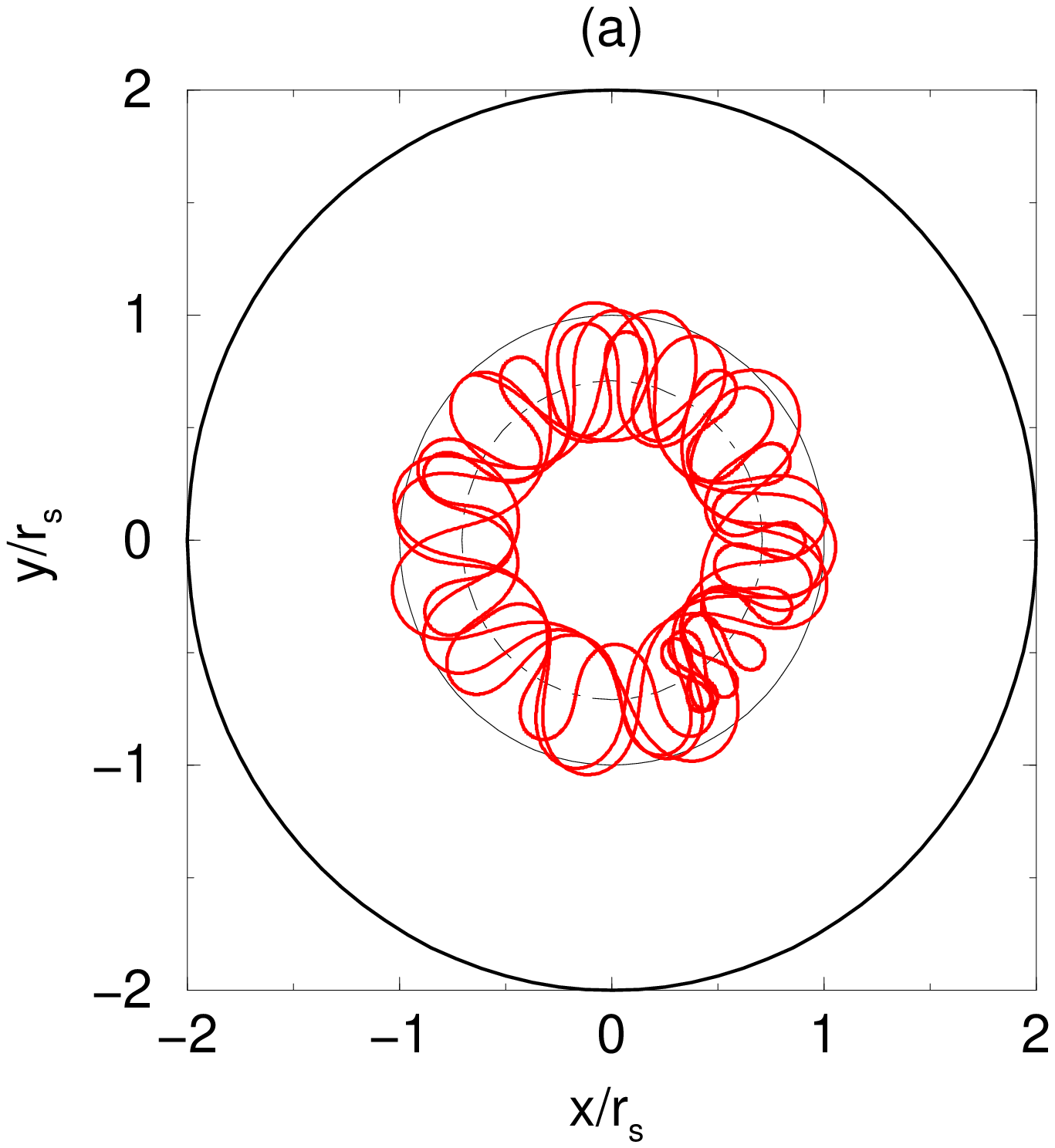}
 \includegraphics[height=7.5cm,angle=+00] {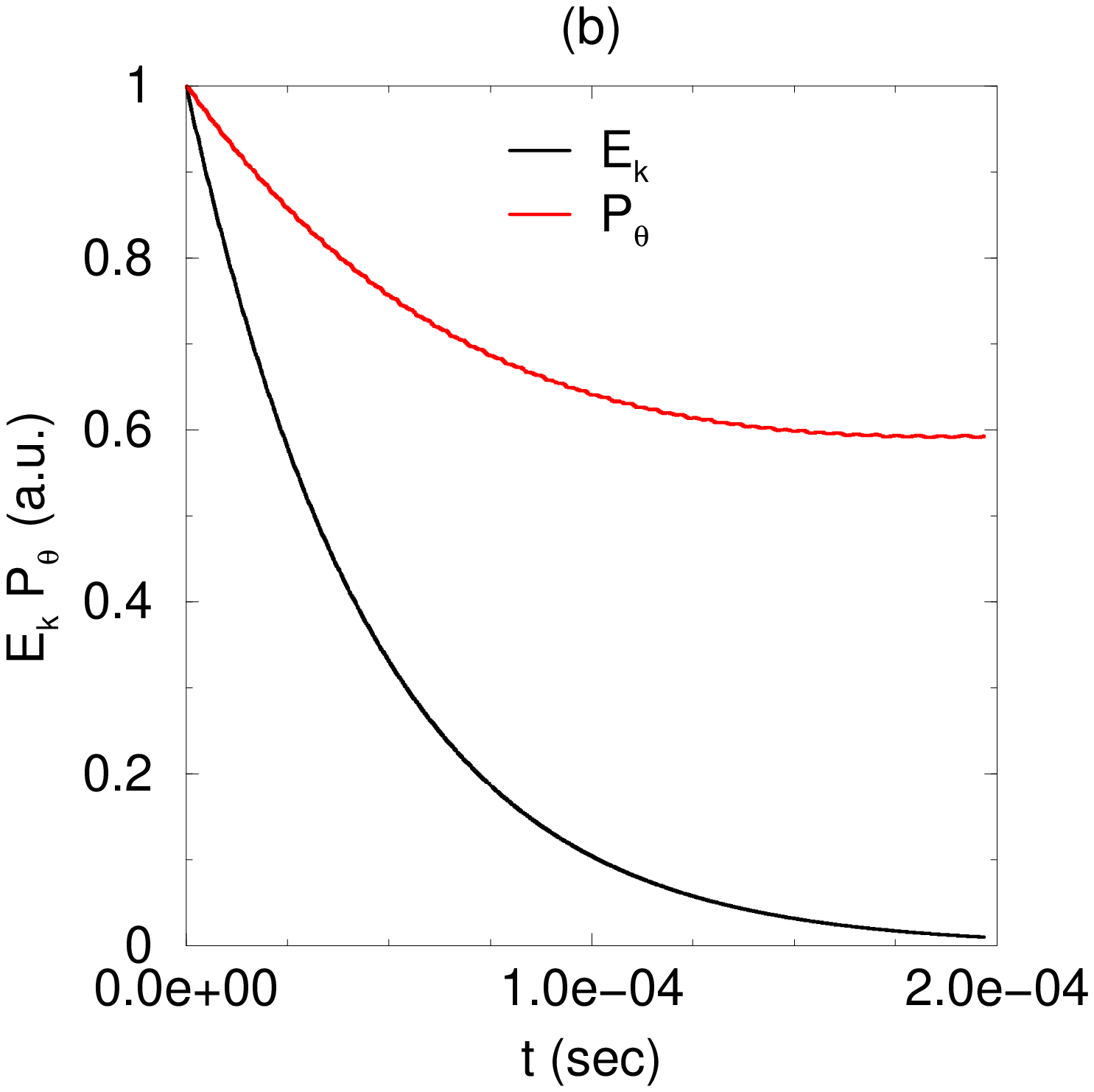}
\caption{(a) Orbital behavior of 100 eV beam ion
in the presence of the collision term. 
The background electron density and temperature is given by
$ 5 \times 10^{19} (m^{-3})$ and $20 (eV)$.
(b) The kinetic energy (black curve) and the angular momentum (red curve)
versus time.}
\label{fig2}
\end{figure}

Figures 2(b) and 3(b) show 
the kinetic energy ($E_k$) and the canonical 
angular momentum ($P_\theta$) versus time.
\begin{figure}
  \centering
 \includegraphics[height=7.5cm,angle=+00] {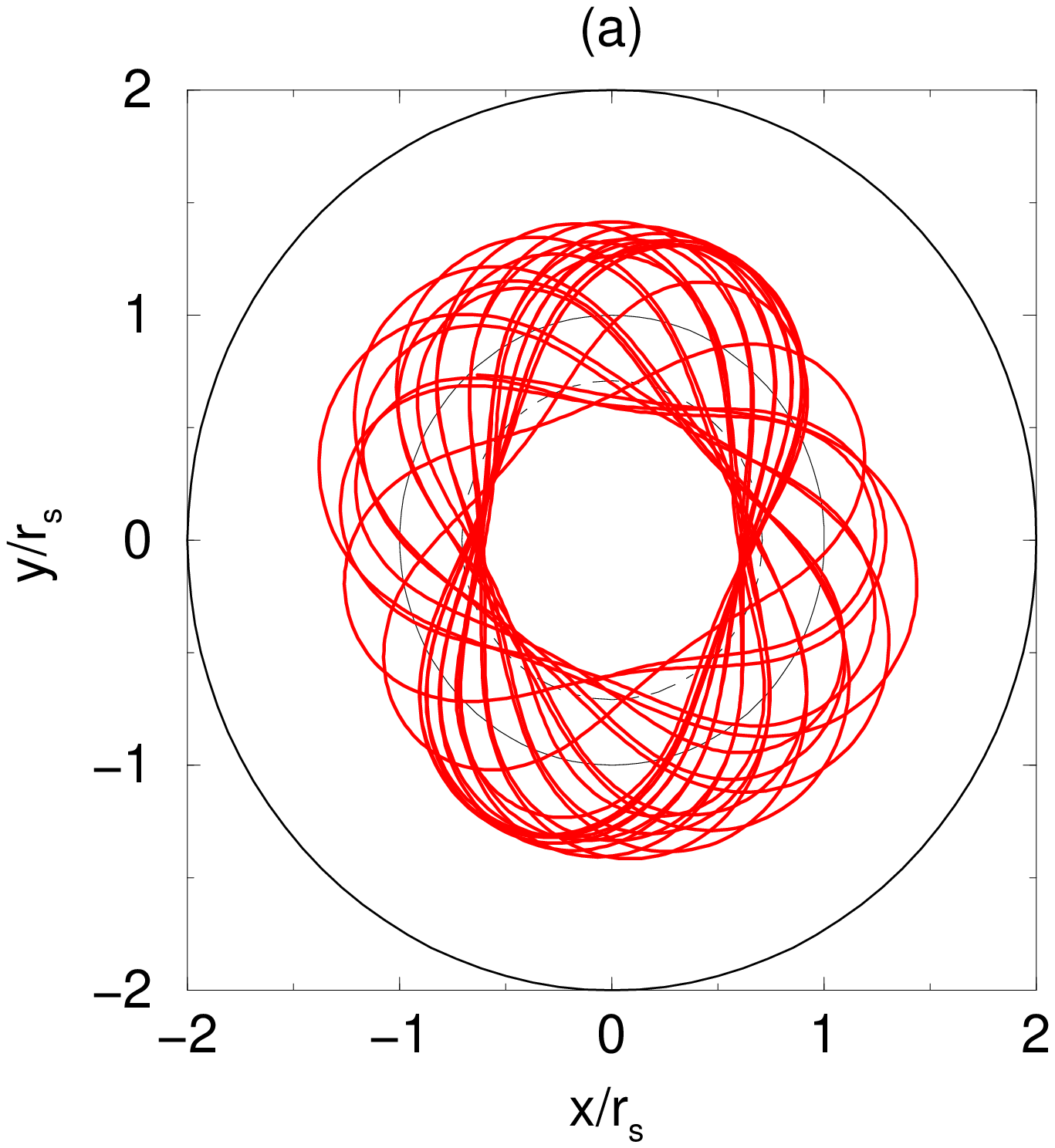}
 \includegraphics[height=7.5cm,angle=+00] {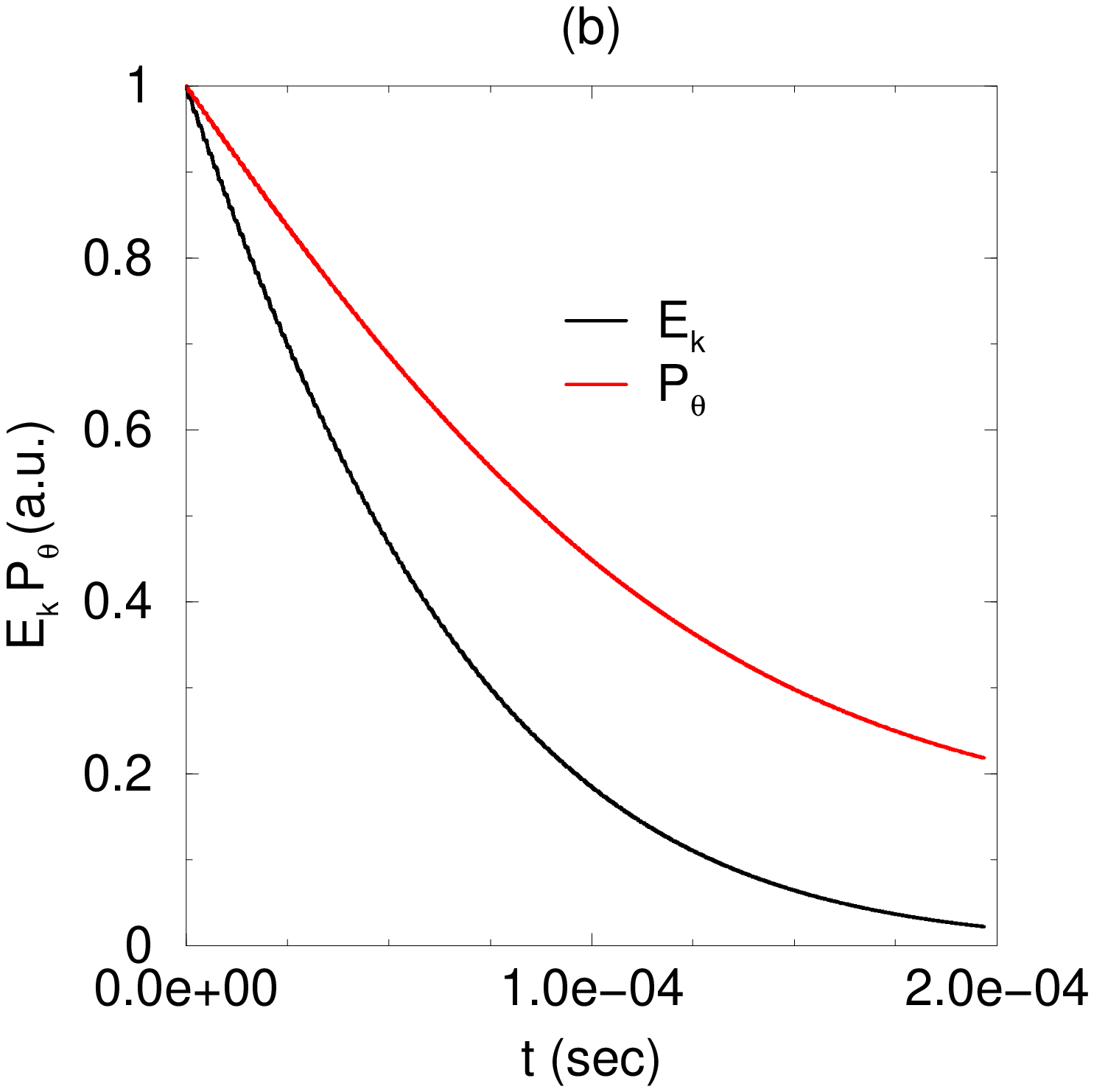}
\caption{(a) Orbital behavior of 2000 eV beam ion
in the presence of the collision term.
The background electron density and temperature is given by
$ 5 \times 10^{19} (m^{-3})$ and $20 (eV)$.
(b) The kinetic energy (black curve) and the angular momentum (red curve)
versus time.}
\label{fig3}
\end{figure}
In Fig.2 and 3, the kinetic energy and the angular momentum relax.
The e-fold times estimated in Fig.2(b) and Fig.3(b) are 
summarized in Table 1 (we take the logarithm; $\tau_{ie}^{sim} $ for the kinetic energy and 
$\tau_{\perp}^{sim}  $ for the angular momentum).

We now compare the numerical relaxation time
[e-fold time estimated from Fig.2(b) and Fig.3(b)]
with a theoretically estimated relaxation time.\cite{miy86,spi62}
Following Ref.\cite{miy86},
the energy relaxation time is given by 
\be
\tau_{ie} = \frac{ \left( 2 \pi \right)^{1/2} 3 \pi \varepsilon_0^2 m m_e}
{ n_e \log \Lambda q^2 q_e^2 }
\left( \frac{T_b}{m} + \frac{T_e}{m_e} \right),
\ee
(which is independent of beam ion temperature unless ${T_b}/{m} \sim {T_e}/{m_e}$)
and the perpendicular momentum relaxation time is given by
\be
\tau_\perp = \frac{ 2 \pi \varepsilon_0^2 m^2 v^3}
{ n_e \log \Lambda q^2 q_e^2 \Phi \left( b_e v \right) },
\ee
respectively. The background plasma
is assumed to be only electrons.
In Eqs.(7) and (8), the charge, the mass, and the temperature of electrons
are given by $q_e$, $m_e$, and $T_e$, respectively.
Here, $b_e = \left(m_e / 2 q_e T_e \right)^{1/2}$.
The relaxation time employing the parameters used
in Fig.2 and Fig.3 are summarized in Table 1.
In Table.1, the energy relaxation time compares favorably with the numerical estimation,  
while the momentum relaxation time differs in particular for the higher energy case.

\begin{table}
\caption{\label{tab:table1} : 
Comparison of relaxation times.}
\begin{tabular}{ccc}
\hline
$T_b$  & $\tau_{ie}$ & $\tau_{ie}^{sim}$  \\
\hline
$100 (eV)$ & $3.9 \times 10^{-5} (s)$ & $4.3 \times 10^{-5} (s)$  \\
$2000 (eV)$ & $3.9 \times 10^{-5} (s)$ & $5.2 \times 10^{-5} (s)$  \\
\hline
$T_b$  & $\tau_\perp$   & $\tau_\perp^{sim}$ \\
\hline
$100 (eV)$ & $1.3 \times 10^{-4}$ (s) & $1.9 \times 10^{-4} (s)$ \\
$2000 (eV)$ & $2.7 \times 10^{-3}$ (s) & $1.3 \times 10^{-4} (s)$ \\
\hline
\end{tabular}
\end{table}

\section{Perturbation method}
\label{s4}
In this section, the perturbation method is introduced. 
Normalizing Eqs.(1) and (2) by the beam ion cyclotron frequency 
$\Omega_c = q_b B_0 / m_b$, and the separatirx radius $r_s$,
we obtain
\be
{d {\bf V} \over     dT} =  {\bf V} \times {\bf B} - \epsilon {\bf F} 
\ee
\be
{d {\bf X} \over     dT} = {\bf V}
\ee
where the frictional force is regarded as a small term employing
\be
\epsilon = {q^4 \log \Lambda n_0 \over 4 \pi \epsilon_0^2 m r_s^3 \Omega_c^4} 
\sum^\star {\Phi_1  (b^\star) \over m_r } \ll 1
\ee
and
\be
{\bf F} = \frac{{\bf V}N(R)}{V^3} 
\sum^\star {\Phi_1 (b^\star V) \over m_r }   
\left( \sum^\star {\Phi_1 (b^\star) \over m_r }\right)^{-1}.
\ee
Expanding ${\bf B} = {\bf B}_0 + \epsilon {\bf B}_1 $ for the rigid rotor profile, we have
\be
{\bf B}_0 = \tanh \left[ \kappa \left( 2 R^2 - 1 \right) \right] {\bf z}
\ee
\be
{\bf B}_1 = 4 \kappa ({\bf X}_0 \cdot {\bf X}_1 ) sech^2 \left[ \kappa \left( 2 R^2 - 1 \right) \right] {\bf z}
\ee
\be
N(R) = sech^2 \left[ \kappa \left( 2 R^2 - 1 \right) \right] 
\ee
Here, the capital letters ($T,{\bf V},R$, and ${\bf X}$) represent the normalized
time, velocity, radius, and position, respectively.

From Eqs.(9) and (10), the lowest order equation is given by
\be
{d {\bf V}_0 \over     dT} =  {\bf V}_0 \times {\bf B}_0
\ee
\be
{d {\bf X}_0 \over     dT} = {\bf V}_0
\ee
and the first order equation in order $\epsilon$ is given by
\be
{d {\bf V}_1 \over     dT} =  {\bf V}_1 \times {\bf B}_0 + {\bf V}_0 \times {\bf B}_1 + {\bf F}
\ee
\be
{d {\bf X}_1 \over     dT} = {\bf V}_1 .
\ee
The solution then is given by the summation ${\bf X} = {\bf X}_0 + \epsilon {\bf X}_1 $,
${\bf V} = {\bf V}_0 + \epsilon {\bf V}_1 $.
The crux in Eqs.(18) and (19) are the 
changes in particle velocity (${\bf V}_1 \times {\bf B}_0$) and particle's displacement 
(${\bf V}_0 \times {\bf B}_1$) both induced by the small friction force (the ${\bf F}$ term). 

The perturbation method is useful since we only need to change the constant $\epsilon$
when the the plasma parameters change, e.g. background densities and temperatures 
(and do not need to recalculate the whole trajectories). 
Figure 4 and 5 show particle trajectories when the perturbation method is employed.
Here, the green curve solution in Fig.5 are obtained by {\it recycling}
${\bf X}_0 $ and $ {\bf X}_1 $ from Fig.4, by simply changing the parameter $\epsilon$.  
In both Figs.4 and 5, the solution from the perturbation method [green curves,
solved Eqs.(16)-(19)]
matches with the direct collision calculations [red curves, solved Eqs.(1) and (2)].
\begin{figure}
 \centering
 \includegraphics[height=7.5cm,angle=+00] {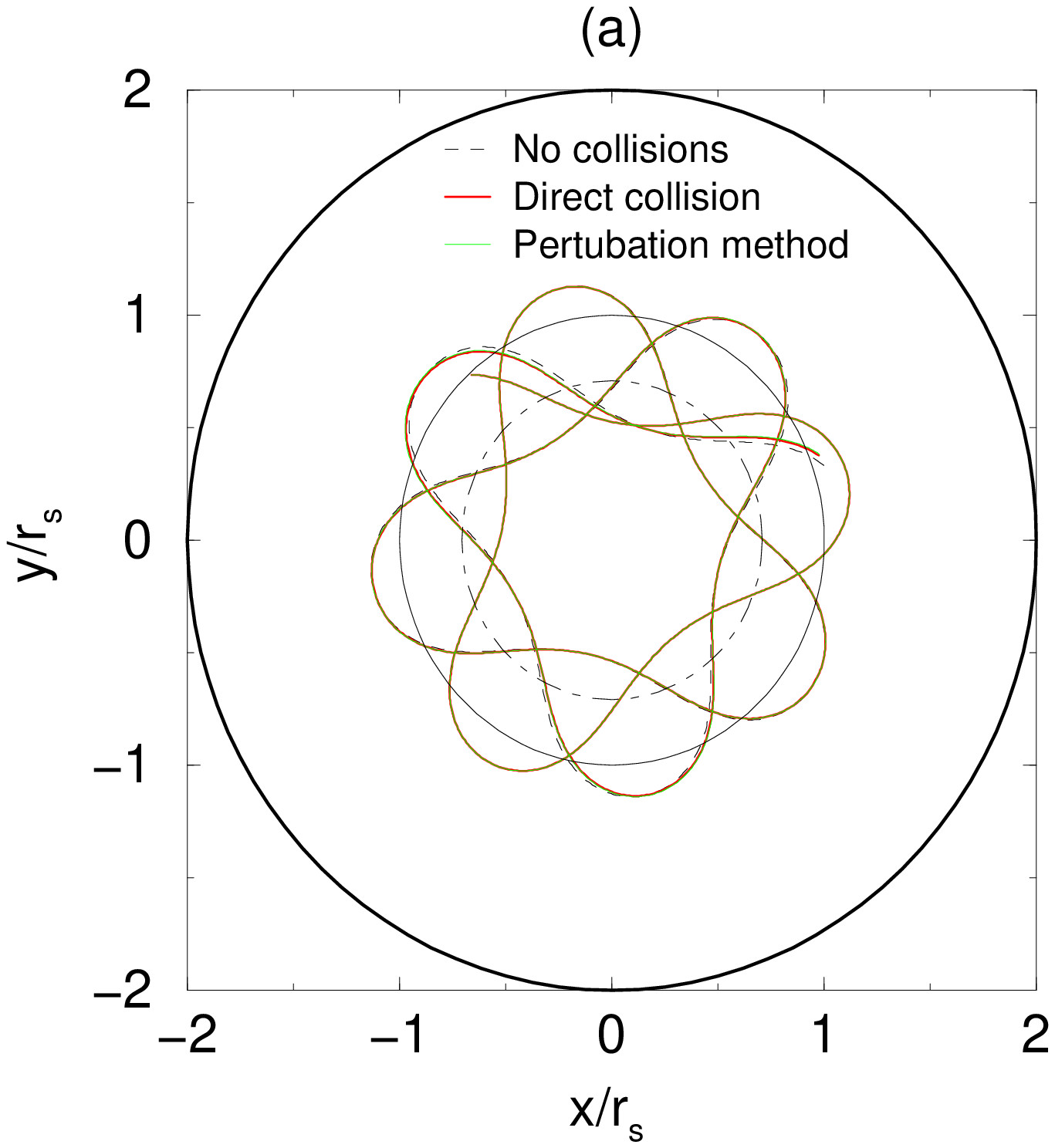}
 \includegraphics[height=7.5cm,angle=+00] {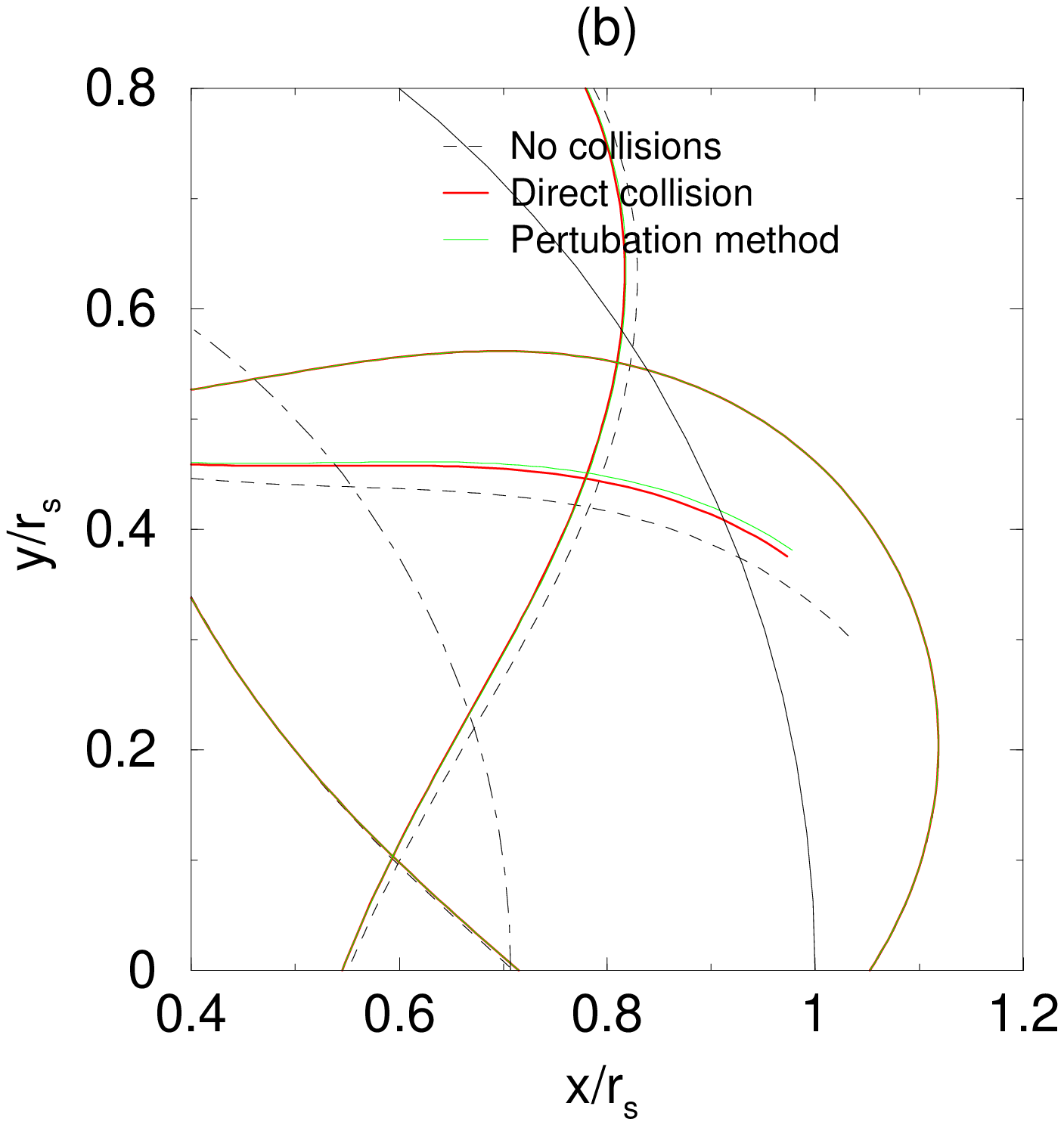}
\caption{
(a) The beam ion orbit with a background plasma
$T_{e} = 50 eV$ and $n_{e} = 1.0 \times 10^{19}$.
The beam ion temperature is $300 (eV)$. 
(b) Expansion of the final stage of Fig.4(a).
The green (red) curve are from the perturbation method (the direct calculation).}
\label{fig4}
\end{figure}

\begin{figure}
 \centering
 \includegraphics[height=7.5cm,angle=+00] {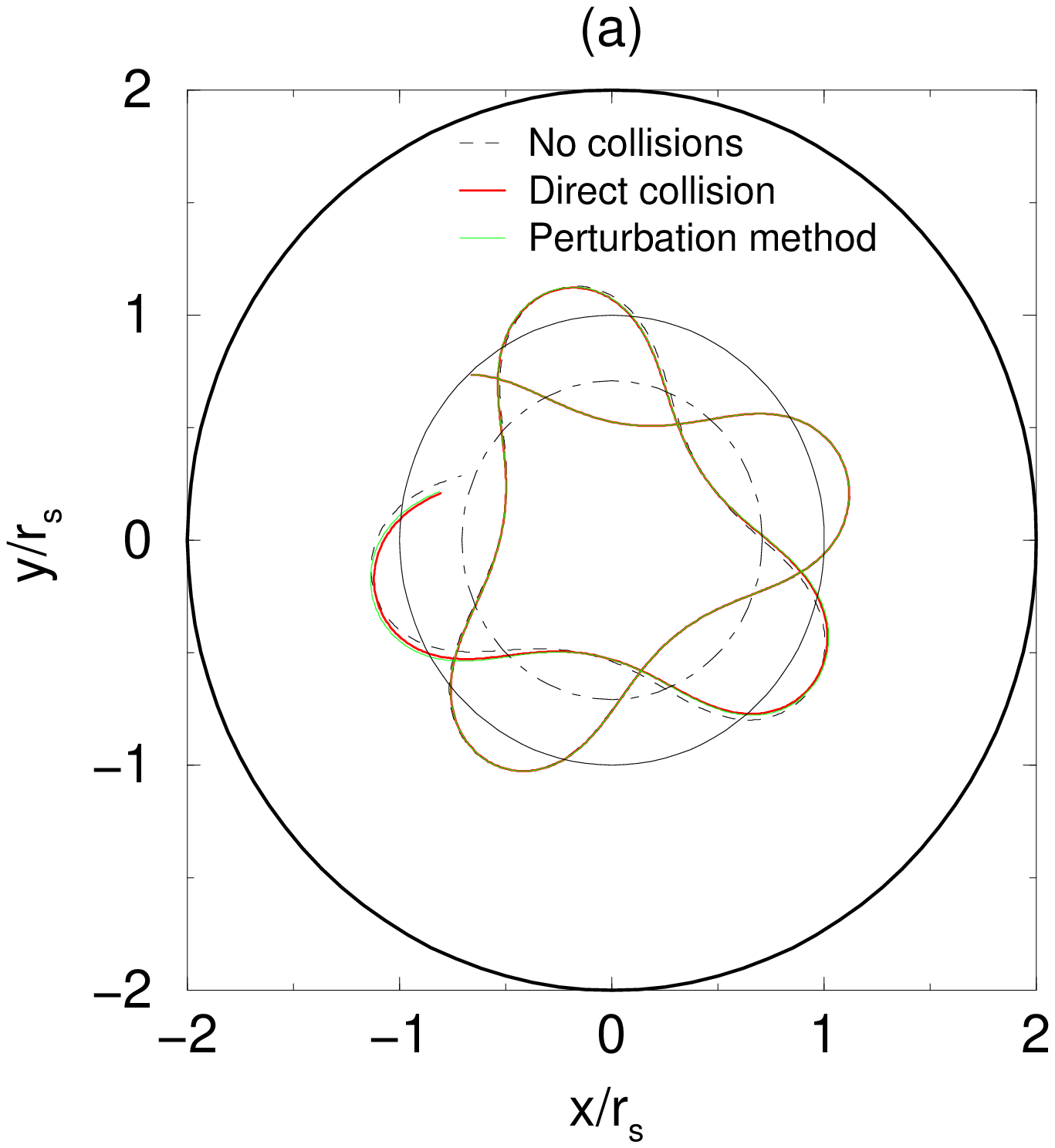}
 \includegraphics[height=7.5cm,angle=+00] {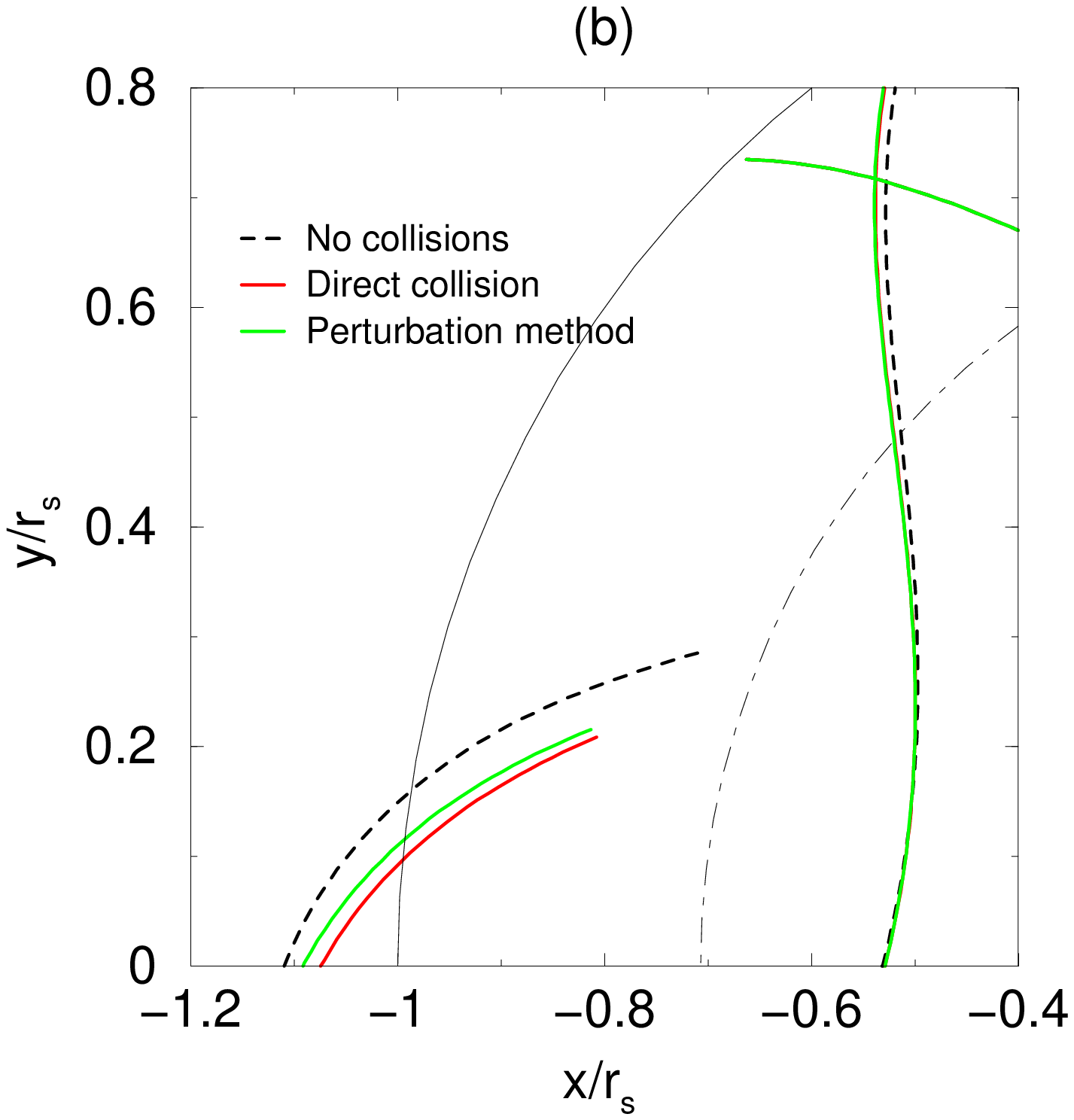}
\caption{
(a) The beam ion orbit with a background plasma
$T_{e} = 50 eV$ and $n_{e} = 5.0 \times 10^{19}$.
The beam ion temperature is $300 (eV)$. 
(b) Expansion of the final stage of Fig.5(a).
Here, the green curve solution in (b) are obtained by {\it recycling}
$X_0 $ and $ X_1 $ from (a), by simply changing the parameter $\epsilon$.
The green (red) curve are from the perturbation method (the direct calculation).}
\label{fig5}
\end{figure}

The lowest order solution is periodic when the magnetic field is axis-symmetric.
Likewise we expect the first order solution to be periodic. 
If the latter is the case, there will be another attractive 
application of the perturbation method. 
By storing the first periodic motion of both the lowest and the higher order solution, 
the algorithm can predict periodic motion in the later phase and thus
can reduce computation time.
As a demonstration, here we take a simplified case where the Lorentz force 
and the centrifugal force are balanced at the initial state;
\be
m r \dot{\theta}^2 = q v B
\ee
[the trajectory will be a perfect
circle in the absence of collisions. See Fig.6(a)]. Figure 6(b) suggests a periodic motion 
of the first order solution from the
perturbation method (time evolution of the
Cartesian coordinates $x_1$ and $y_1$ are plotted).
\begin{figure}
 \centering
 \includegraphics[height=7.5cm,angle=+00] {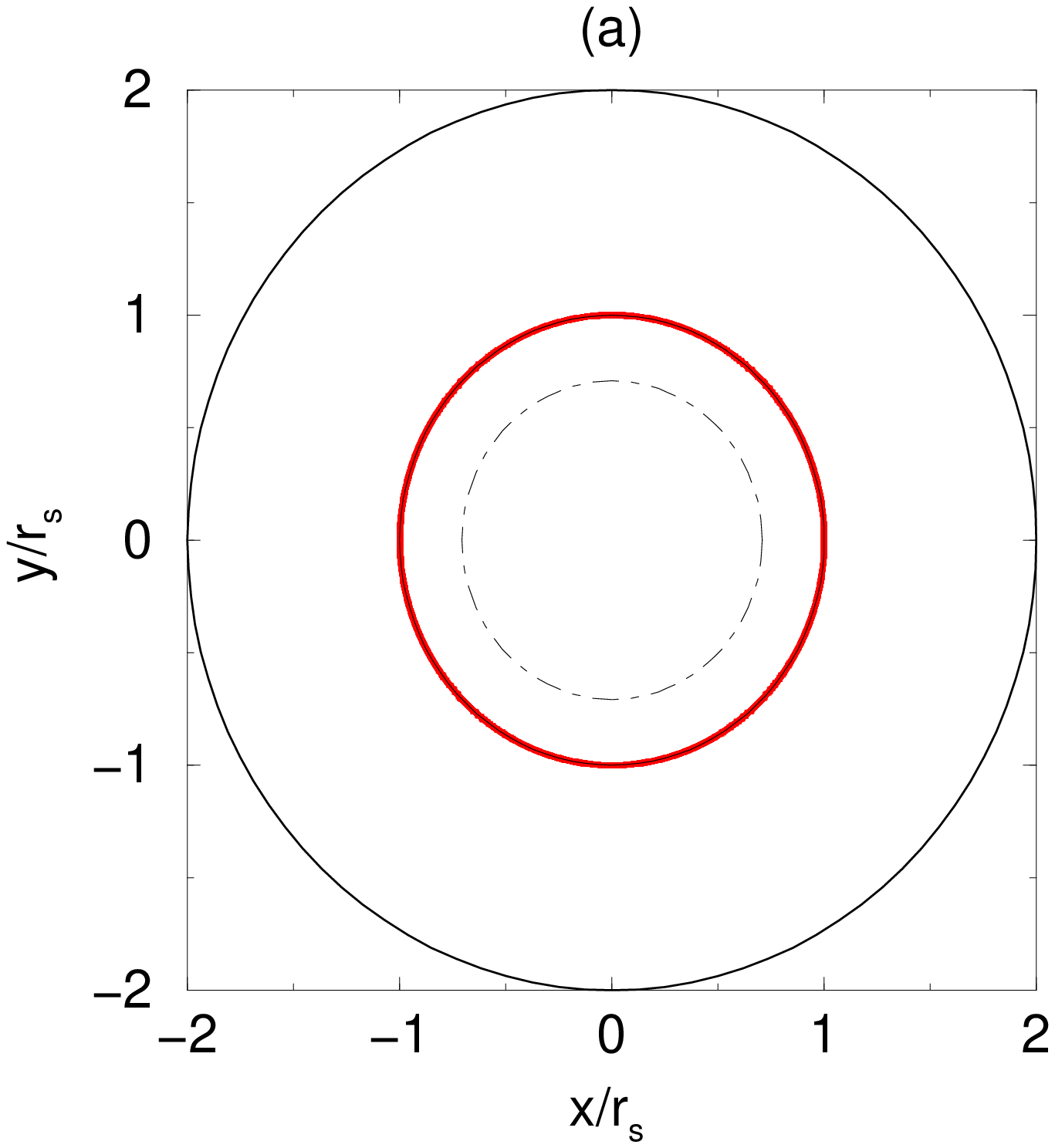}
 \includegraphics[height=7.5cm,angle=+00] {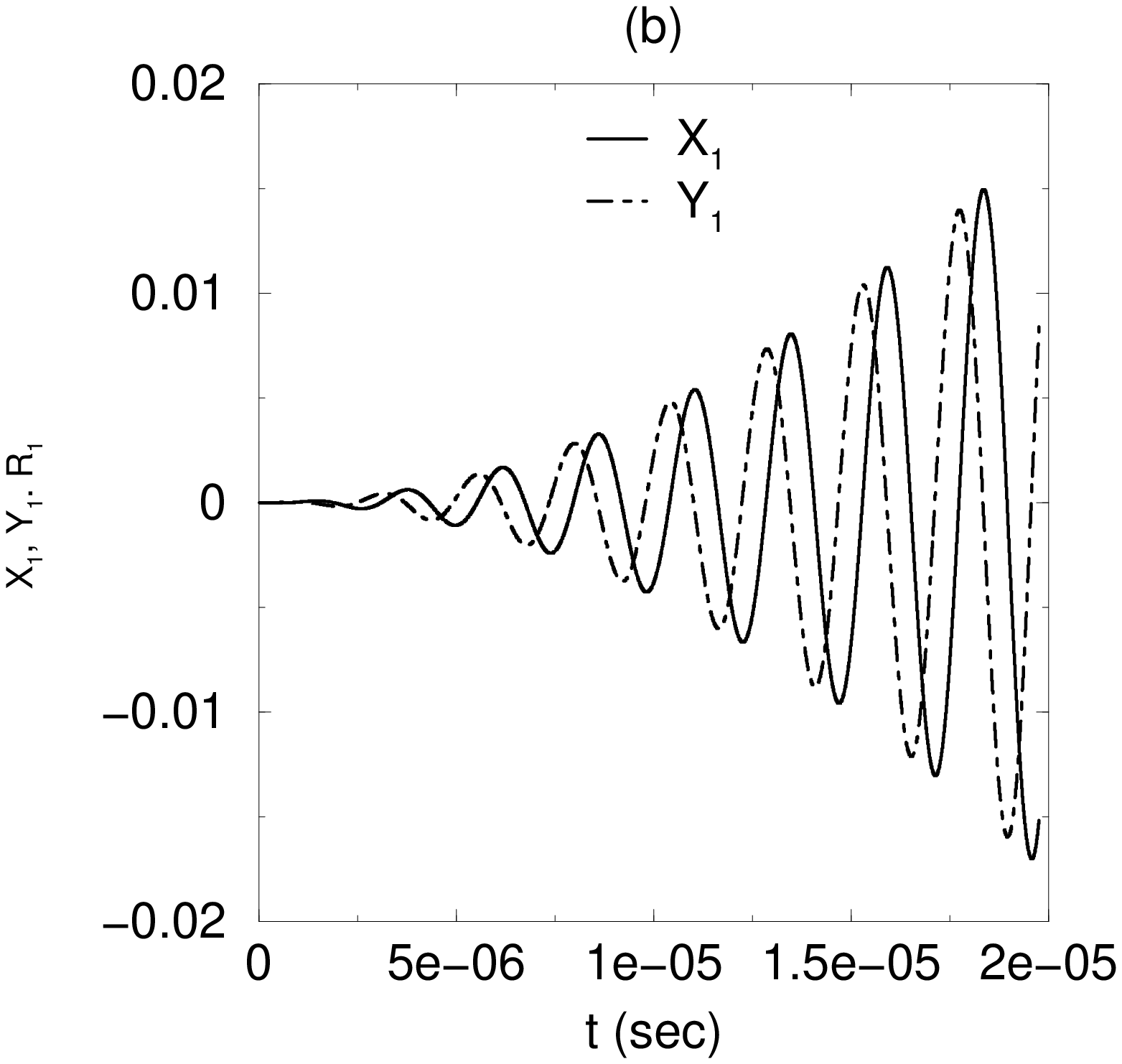}
\caption{(a) The beam ion orbit (red circle) when the Lorentz force
and the centrifugal force are balanced at the initial state.
The initial position is at $x=0$ and $y=r_s$.
(b) Time evolution of the perturbed quantities
$x_1$ (solid), $y_1$ (dashed).}
\label{fig6}
\end{figure}

\section{Summary}
\label{s5}
An orbit following code is developed to calculate 
ion beam trajectories in magnetized plasmas.
The equation of motion is solved 
incorporating the Lorentz force term and Coulomb collisional relaxation term.
Conservation of energy and angular momentum is confirmed in
the absence of collisions.
With the collisions, it is shown that the 
energy relaxation time compares favorably
with the theoretical prediction.\cite{miy86}

Furthermore,
a new algorithm to calculate ion beam trajectories is reported. 
We have applied perturbation method regarding the collisional
term small compared to the zeroth order Lorentz force term.
The two numerical solutions from the perturbation
method and the direct collisional calculation matched.
The perturbation method is useful since we only need to change the 
perturbation parameter to recalculate the trajectories, when
the background parameters change. 
In general, the algorithm can be applied 
to periodic motion under perturbative frictional forces, 
such as guiding center trajectories\cite{mor66,lit83,nis97,nis08} or satellite motion. 
We have also suggested a reduction in computation time
by capturing the periodic motion. More detailed analysis
will be our future work.

The work is initiated as a diploma thesis at 
Osaka University during the years 1990-1991.\cite{nis91}
The author would like to
thank Dr. T.~Ishimura and Dr. S.~Okada for useful discussions.
A part of this work is supported by National Cheng Kung University
Top University Project.
The author would like to
thank Dr.~C.~Z.~Cheng and Dr.K.~C.~Shaing. \\

\appendix{\bf{Appendix}}

We recapitulate Coulomb collision processes presented in Ref.\cite{miy86}.
Assuming the background plasma is Maxwellian,
change in the momentum for the beam ion within time $dt$ is given by
\be
\left< {d {\bf p} \over dt} \right> = 
- {q^2 {\bf v} \over 4 \pi \varepsilon_0^2 v^3} \sum^\star {\log \Lambda (q^\star)^2 \over m_r} n^\star 
\Phi_1 (b^\star v) 
\ee
here, $q$ ($q^\star$) and $m$ ($m^\star$) are the charge and the mass of 
the beam ions (background plasma species), $m_r = m m^\star / ( m + m^\star ) $ 
is the reduced mass, $n^\star$ is the background plasma density.
Here, $\sum^\star $ signifies summation over species.
The Coulomb logarithm is given by
\be
\log \Lambda \simeq 7 + \log \left[ \left( {T_e \over e} \right)^{3/2} / 
\left( {n_e \over 10^{20}} \right)^{1/2} \right] 
\ee
where the electron temperature is in the unit of $e V$,
and the electron density is in the unit of $m^{-3}$.

Letting
$x = b^\star v  = \left( m^\star / 2 e T^\star \right)^{1/2} 
\left( 2 e T / m \right)^{1/2} = \left( m^\star T / m T^\star \right)^{1/2}$, 
the function $\Phi (x)$ is given by
\be
\Phi (x) = {2 \over \sqrt{\pi}}  \int_0^x \exp{\left( - \xi^2 \right)} d\xi 
\ee
\be
\Phi_1 (x) = \Phi (x) - {2 x \over \sqrt{\pi}}  \exp{\left( - x^2 \right)} .
\ee
In Eq.(21), the contribution from electrons is much larger
than the ions because of the $1/m_r$ factor.
The relaxation of high energy ions is namely due to the collision
with the electrons. One can then employ the form
\be
\Phi_1 (x) = {4 x^3 \over 3 \sqrt{\pi}}.
\ee
at the $x \ll 1$ limit [the integration form of Eqs.(23) and (24)
are employed in the computation]. \\

\appendix{\bf{References}}

\end{document}